\documentstyle[preprint,aps]{revtex}
\begin{document}
\draft

\author{A. R. Denton* and J. Hafner}

\address
{Institut f\"ur Theoretische Physik, Technische Universit\"at Wien\\
Wiedner Hauptstra{\ss}e 8-10, A-1040 Wien, Austria}

\title{Thermodynamically Stable One-Component Metallic Quasicrystals}

\date{\today}
\maketitle

\begin{abstract}
Classical density-functional theory is employed to study 
finite-temperature trends in the relative stabilities of 
one-component quasicrystals interacting via 
effective metallic pair potentials
derived from pseudopotential theory.  
Comparing the free energies of several periodic crystals and 
rational approximant models of quasicrystals over a range of 
pseudopotential parameters, {\it thermodynamically} stable quasicrystals
are predicted for parameters approaching the limits of mechanical stability 
of the crystalline structures.  
The results support and significantly extend conclusions of 
previous ground-state lattice-sum studies.
\end{abstract}

\bigskip
\bigskip
\pacs{PACS numbers: 61.44.+p, 64.70.Dv, 61.25.Mv}

\newpage




Since the landmark discovery~\cite{Shechtman} of 
long-range icosahedral quasiperiodic ordering in Al-Mn alloys, 
quasicrystals have been produced in a rich variety of systems~\cite{QC,Tsai}.
All known quasicrystals, however, are alloys of at least two 
metallic elements.  
A fundamental question has therefore naturally arisen:  can one-component 
quasicrystals be thermodynamically stable?
A ground-state icosahedral phase has been predicted 
for an idealized square-well pair potential system~\cite{Jaric1}, 
although the stability range is confined to a narrow range of 
well widths and pressures.
More recently, extensive lattice-sum potential energy 
calculations~\cite{latsum} for systems interacting 
via effective metallic pair potentials 
have predicted energetically stable 
ground-state one-component quasicrystals, albeit within a 
restricted range of pair potential parameters 
having no counterparts in the Periodic Table.  
An alternative approach, especially suited to finite temperatures, 
is classical density-functional (DF) theory~\cite{DF1,DF2}, which determines 
the free energy of a given solid as the variational minimum 
with respect to density of an approximate free energy functional.
Fundamental applications have already predicted purely entropic 
hard-sphere quasicrystals to be either metastable~\cite{MA} 
or mechanically unstable~\cite{Jaric2}, indicating that 
ordinary entropy alone is not sufficient 
to stabilize quasiperiodicity. 
The question remains whether at finite temperatures 
longer-range interactions may conspire to
stabilize one-component quasicrystals.
In this Letter we directly address this question by extending 
classical DF methods to simple metals interacting via 
effective pair potentials derived from pseudopotential theory.
Taking as a structural model of quasicrystals
a certain class of rational approximants, we predict, 
over a limited range of pseudopotential parameters, 
thermodynamically stable one-component quasicrystals, 
stabilized largely by medium- and long-range interactions.


Interactions between ions in a metal are complicated by the presence of
conduction electrons and are often best determined by electronic DF methods.
However, in cases where the conduction electrons may be considered 
nearly free, as in the simple metals, it proves possible to replace 
the strong electron-ion interaction by a much weaker {\it pseudopotential}, 
which is amenable to perturbation theory~\cite{HM}. 
The pseudopotential approach~\cite{pseud}
essentially reduces the two-component system of electrons and ions 
to an effective one-component system of pseudoatoms, interacting via
an effective pair potential $\phi(r)$ that depends on 
the pseudopotential and the density-dependent dielectric function.  
Here we adopt the popular empty-core pseudopotential~\cite{Ashcroft}, 
parametrized by valence $Z$ and core radius $r_c$, and the
Ichimaru-Utsumi dielectric function~\cite{IU},  
which satisfies important self-consistency conditions.
For the simple metals, $\phi(r)$ is typically characterised by 
a steeply repulsive short-range core, a first minimum of varying range
and depth, and a relatively weak oscillating tail. 
Of particular relevance here is that
with increasing $Z$ or decreasing $r_c$ the (Friedel) oscillations
weaken in amplitude and shorten in wavelength.  Correspondingly
the core shrinks and, as the oscillations move under the core, 
a repulsive shoulder may develop.


We now proceed to take $\phi(r)$ as input to a DF theory
for a purely classical one-component system 
of pair-wise interacting pseudoatoms. 
The DF approach~\cite{DF1,DF2} defines a functional $F[\rho]$ 
of the nonuniform one-particle number density $\rho({\bf r})$ 
that satisfies a variational principle,  
whereby $F[\rho]$ is minimized (at fixed average density) 
by the equilibrium $\rho({\bf r})$, 
its minimum value equaling the Helmholtz free energy.
In practice, $F[\rho]$ is separated into an exactly known ideal-gas 
contribution $F_{id}$, the free energy of the nonuniform system 
in the absence of interactions, and an excess contribution $F_{ex}$, 
depending entirely upon internal interactions. 
Note that we ignore here a sizable volume-dependent, but
structure-independent, contribution~\cite{HM}, 
which has no bearing on structural stabilities 
at fixed average density.
The form of $\phi(r)$ for the simple metals
naturally suggests decomposition into a 
short-range reference potential 
$\phi_o(r)$ and a perturbation potential 
$\phi_p(r)=\phi(r)-\phi_o(r)$.
Such an approach has already been applied with success to 
Lennard-Jones solids~\cite{CA2,Mederos}
as well as to metallic solids~\cite{DHK}.
To first order in the perturbation potential~\cite{HM},
\begin{equation}
F[\rho] \simeq F_o[\rho] + {1\over{2}}\int d{\bf r}\int d{\bf r}'
\rho({\bf r})\rho({\bf r}')g_o({\bf r},{\bf r}')
\phi_p(|{\bf r}-{\bf r}'|),
\label{pert}
\end{equation}
where $F_o[\rho]$ is the reference free energy functional
and $g_o({\bf r},{\bf r}')$ the reference pair distribution function.
Following Weeks, Chandler, and Andersen (WCA)~\cite{WCA},  
we split $\phi(r)$ at its first minimum and map
the steeply repulsive reference system onto a system of 
hard spheres of effective diameter $d$, determined at temperature $T$
by the Barker-Henderson prescription~\cite{HM}
$d = \int_0^{\infty}dr\{1-\exp[-\phi_o(r)/k_BT]\}$.
The free energy of the liquid is calculated
via the uniform limit of Eq.~(\ref{pert}), 
using the accurate Carnahan-Starling $f_{HS}(\rho)$
and Verlet-Weis $g_{HS}(r)$~\cite{HM}.

The solid reference free energy is now approximated by the 
modified weighted-density approximation (MWDA)~\cite{MWDA,CA1}, 
which is known to give an accurate description of the HS system~\cite{MWDA}.
This maps the excess free energy per particle of the solid 
onto that of the corresponding {\it uniform} fluid $f_{HS}$, according to
\begin{equation}
F_{ex}^{MWDA}[\rho]/N = f_{HS}(\hat\rho),
\label{MWDA}
\end{equation}
where the effective (or weighted) density
\begin{equation}
\hat\rho \equiv \frac{1}{N} \int d{\bf r} \int d{\bf r}'
\rho({\bf r})\rho({\bf r}')w(|{\bf r}-{\bf r}'|;\hat\rho)
\label{rhohat1}
\end{equation}
is a self-consistently determined weighted average of $\rho({\bf r})$.  
The weight function $w(r)$ is specified by normalization 
and by the requirement that $F_{ex}^{MWDA}[\rho]$ generate the 
exact two-particle (Ornstein-Zernike) direct correlation function 
$c(r)$ in the uniform limit.  
This leads to an analytic relation~\cite{MWDA} 
between $w(r)$ and the fluid functions $f_{HS}$ and $c(r)$, computed here
using the solution of the Percus-Yevick integral equation 
for hard spheres~\cite{HM}.
The perturbation free energy depends on the 
pair distribution function of the reference HS solid 
$g_{HS}({\bf r},{\bf r}';d)$, which we approximate in mean-field fashion 
by the Heaviside unit step function $u(|{\bf r}-{\bf r}'|-d)$.
This is justified by the fact that in the ordered solid
most of the structure of the two-particle density 
is contained already at the level of the one-particle density, rendering
$g_{HS}({\bf r},{\bf r}';d)$ a relatively structureless function~\cite{CA1}.

Practical calculation of $F[\rho]$
requires specification of the solid structure. 
Here we consider fcc, hcp, and bcc elementary crystals 
and a quasicrystalline structure, modelled by a Penrose tiling 
constructed by projecting a six-dimensional hypercubic lattice 
onto the three-dimensional physical space.
In particular, we consider a hierarchy of 
rational approximants~\cite{QC,RA}
based on the ``unit-sphere packing" model of Henley~\cite{Henley}. 
These are periodic structures obtained by replacing, in the three-dimensional 
perpendicular space, the golden mean $\tau$ by a
rational number $\tau_n=F_{n+1}/F_n$, where $F_n$ is a term
in the Fibonacci sequence.
The first four approximants, denoted by 1/1, 2/1, 3/2, and 5/3, have
unit cells of $20$, $108$, $452$, and $1904$ atoms, respectively, 
and corresponding maximum HS packing fractions of $0.5020$, $0.6400$, 
$0.6323$, and $0.6287$, compared with $0.6288$ in the quasiperiodic limit 
and $0.7405$ for the close-packed fcc and hcp crystals.
Also to be specified is the atomic density distribution about the 
lattice sites of the solid, for which we adopt the widely used
Gaussian ansatz. 
This places at each site ${\bf R}$ a normalized isotropic Gaussian, 
such that
\begin{equation}
\rho({\bf r})=\left(\frac{\alpha}{\pi}\right)^{3/2}\sum_{\bf R}
\exp(-\alpha|{\bf r}-{\bf R}|^2),
\label{Gauss}
\end{equation}
where the single parameter $\alpha$ determines the width 
of the distribution. 
The Gaussian ansatz is known~\cite{DF2,YA_OLW} to reasonably describe
the density distribution of close-packed crystals near melting
and also should be a good approximation for elastically isotropic
quasicrystals.


At fixed $T$, atomic volume $\Omega$, $Z$, and $r_c$, 
the free energy is computed 
from Eqs.~(\ref{pert})-(\ref{Gauss}) for a given solid structure
by minimizing $F[\rho]$ with respect to $\alpha$ 
(and the $c/a$ ratio for the hcp crystal).  A minimum 
at $\alpha{\neq}0$ implies mechanical stability. 
Finally, comparing free energies of
different solid structures and the liquid, 
the thermodynamically stable structure 
is determined as that having the lowest free energy.
As an essential test of the theory, we have applied it to the 
HS reference system alone~\cite{DH}.
Briefly, the quasicrystals are predicted to be mechanically stable 
over a wide range of densities, but always metastable relative to both  
the fluid and the crystals.  Their free energies decrease
in order of increasing maximum packing fraction, 
confirming the decisive importance of packing efficiency in the HS system.
Furthermore, their Lindemann ratios (root-mean-square atomic displacement 
over nearest-neighbour distance) are significantly smaller than those of 
the crystals, reflecting a higher degree of atomic localization.
As a further test of the theory, and of the pseudopotential approach, 
we have examined the freezing transitions of two simple metallic elements, 
Mg ($Z=2$) and Al ($Z=3$), interacting via effective 
pair potentials~\cite{DHK}.
In both cases, the theory correctly predicts 
the stable equilibrium structure (hcp for Mg, fcc for Al)
and structural energy differences in reasonable agreement with experiment.  

We now turn to a broad survey of pseudopotential parameters 
for the simple metals.
Table~I summarizes our results for the thermodynamic state defined by 
$T=500 K$ and ${\Omega}=150 a_o^3$ ($a_o=$ Bohr radius)
over the ranges $2.0\leq Z\leq 3.4$ and $0.400\leq r_c/r_s\leq 0.575$, 
where $r_s\equiv (3{\Omega}/4{\pi}Z)^{1/3}$ is the electron-sphere radius.
Note that we treat $Z$ here as a continuous parameter to better study 
trends in relative stabilities.
For orientation, we note that near-equilibrium hcp-Mg occurs at 
$r_c/r_s=0.501$, while expanded fcc-Al would occur at $r_c/r_s=0.486$
were it not unstable at such low densities (see below).
By default, the liquid is considered the stable phase when 
the effective HS packing fraction $\eta\equiv \frac{\pi}{6}\rho d^3$
is so low that no solid structure is stable against atomic displacements.  
At sufficiently high $\eta$, where repulsive short-range interactions 
favour packing efficiency, the close-packed crystals are
thermodynamically stable, the minimizing hcp $c/a$ ratio 
lying close to the ideal ratio $\sqrt{8/3}$.
The more open bcc structure is always at best metastable.
Remarkably though, in a narrow band of parameter space
corresponding to intermediate $\eta$, 
the quasicrystal structures are predicted to be 
{\it thermodynamically} stable. 
Although the relevant parameter range contains none of 
the elements from the Periodic Table, it does include the 
virtual-crystal parameters (average $Z$, $r_c$, and $\Omega$) 
characteristic of some of the known multi-component quasicrystals.

We now analyse the physical reasons for quasicrystal stability. 
When $\eta$ is so low that the free energy functional has no 
variational minimum, the solid exhibits a {\it phonon instability}.
Such unstable regions in the parameter space of Table~I can be understood 
by noting that the effective HS diameter in general decreases 
with increasing $Z$ (across rows) and with decreasing $r_c$ (down columns).
Correspondingly, the solid becomes more loosely packed and the
density distributions broader, as reflected by increasing $L$
for a given structure.  
Ultimately, when $L>0.10-0.15$, the solid loses its stability against 
vibrational atomic displacements (phonons).  
As in the HS system, the quasicrystals
have consistently smaller $L$ than the crystals, 
implying greater vibrational stiffness, which tends to enhance 
their mechanical stability.
This partly explains their appearance 
near the diagonal of Table~I, where the crystals 
lose mechanical stability, as well as the absence of any stable structure 
towards the lower-right corner.
Where no stable structure is indicated, $F[\rho]$ cannot be evaluated 
within the theory for any of the structures considered.
The reason is that at sufficiently high $Z$ and short $r_c$, 
when $\phi(r)$ develops a repulsive shoulder, $d$ may be anomalously large. 
When $d$ is so large that the maximum $\eta$
for a given structure is exceeded, 
the repulsive cores of nearest-neighbour atoms overlap 
at equilibrium separation. 
This implies a {\it packing instability}, which is a second reason 
for loss of structural stability upon approaching 
the lower-right corner of Table~I.
In this parameter region, analysis of structural trends across 
the Periodic Table~\cite{pseud} and lattice-sum calculations~\cite{latsum} 
indicate the stability of more open covalent structures, 
which are, however, outside the scope of the present theory. 

Phonon and packing instabilities account for loss of mechanical stability.
Insight into the source of {\it thermodynamic} stability is gained by 
examining separate contributions to the total free energy. 
The reference free energy $F_o$ combines the entropy and the part 
of the internal energy associated with short-range interactions, 
while the perturbation free energy $F_p$ is the remaining part 
of the internal energy deriving from longer-range interactions.
Figure~1 compares $F_o$ and $F_p$ as a function of $Z$ 
for the various structures at fixed $r_c/r_s$ and $\Omega$.  
As expected, $F_o$ consistently favours the compact crystals. 
More revealing is that as $Z$ increases, $F_p$ increasingly favours 
the more open quasicrystals.  This can be understood 
by noting that with increasing $Z$, as the Friedel oscillations move under 
the repulsive core, the first minimum of $\phi(r)$ shifts from 
the nearest-neighbour distance of the highly-coordinated fcc and hcp 
structures to shorter distances more commensurate with the 
lower-coordinated rational approximant structures.
Thus the oscillating tail of $\phi(r)$ clearly emerges as
the source of thermodynamic stability of simple metallic quasicrystals.


Summarizing, using a practical DF-perturbation theory,  
we have calculated the free energies
of simple metallic one-component crystals and quasicrystals, 
interacting via effective pair potentials.
Separating the full pair potential into a steeply repulsive 
short-range core and an oscillating tail, the theory 
minimizes an approximate free energy functional 
with respect to the density distribution.
By surveying a range of pair potential parameters,  
we have identified general trends in relative stabilities.
With increasing valence and decreasing core radius, 
as the close-packed crystals lose mechanical stability, 
thermodynamically stable quasicrystals are predicted to emerge.
Besides extending to finite temperatures the intriguing predictions 
of previous ground-state lattice-sum studies~\cite{latsum}, 
our approach yields new physical insight by linking
quasicrystalline stability to enhanced vibrational stiffness 
and a competition between short- and medium-long-range interactions. 
Future application to mixtures may help
to further explain the stability of real quasicrystalline alloys.  

\acknowledgements
\noindent
We thank G. Kahl, M. Kraj\^c\'i, and M. Windisch 
for numerous helpful discussions.
This work was supported by the Fonds zur F\"orderung der wissenschaftlichen 
Forschung (Austrian Science Foundation) to whom one of us (ARD) is grateful 
for a Lise-Meitner Fellowship.

\bigskip
\noindent
*~Present address: Institut f\"ur Festk\"orperforschung, 
Forschungszentrum J\"ulich GmbH, 
D-52425 J\"ulich, Germany (e-mail: a.denton@kfa-juelich.de)






\newpage

\unitlength1mm

\begin{figure}
\noindent
\caption[]{
(a) Reference hard-sphere free energy, and (b) perturbation
free energy, for designated solid structures, as a function of valence $Z$
at fixed core radius $r_c/r_s=0.550$, atomic volume ${\Omega}=150 a_o^3$, 
and temperature $T=500 K$.  
}
\label{FIG1}
\end{figure}


%
%
%

\begin{table}
\bigskip
\bigskip
\noindent
\caption{
Thermodynamically stable structure, Lindemann ratio, 
and effective hard-sphere diameter (in Bohr radii), 
at fixed atomic volume ${\Omega}=150 a_o^3$ and temperature $T=500 K$,  
over a range of valence $Z$ and core radius $r_c$, as a ratio of 
electron-sphere radius $r_s$.
(Parentheses indicate metastability with respect to the liquid phase.)
}
\label{TAB1}
\bigskip
\bigskip
\bigskip
\begin{tabular}{cc|cccccccc}
 & & & & &$Z$& & & & \\
\hline
$r_c/r_s$& &2.0&2.2&2.4&2.6&2.8&3.0&3.2&3.4\\
\hline
\hline
0.575& &hcp&hcp&fcc&fcc&fcc&5/3&5/3&5/3\\
 & &0.0258&0.0285&0.0395&0.0450&0.0507&0.0453&0.0526&0.0631\\
 & &5.794&5.692&5.601&5.519&5.445&5.378&5.317&5.262\\
\hline
0.550& &hcp&hcp&fcc&fcc&3/2&5/3&5/3&5/3\\
 & &0.0315&0.0356&0.0495&0.0554&0.0516&0.0581&0.0690&0.0876\\
 & &5.677&5.584&5.501&5.426&5.359&5.299&5.245&5.198\\
\hline
0.525& &hcp&fcc&fcc&fcc&5/3&5/3&(5/3)&liquid\\
 & &0.0431&0.0558&0.0620&0.0706&0.0653&0.0771&0.0967&--\\
 & &5.553&5.470&5.395&5.329&5.271&5.219&5.174&5.135\\
\hline
0.500& &hcp&fcc&fcc&(2/1)&(5/3)&(5/3)&liquid&liquid\\
 & &0.0625&0.0708&0.0802&0.0789&0.0878&0.109&--&--\\
 & &5.422&5.348&5.283&5.228&5.179&5.138&5.104&5.078\\
\hline
0.475& &hcp&fcc&(fcc)&(2/1)&(3/2)&liquid&liquid&liquid\\
 & &0.0822&0.0931&0.113&0.110&0.135&--&--&--\\
 & &5.282& 5.219&5.166&5.121&5.085&5.058&5.042&5.042\\
\hline
0.450& &(hcp)&(fcc)&(2/1)&liquid&liquid&liquid&--&--\\
 & &0.111&0.137&0.134&--&--&--&--&--\\
 & &5.136&5.085&5.044&5.013&4.993&4.989&6.964&6.955\\
\hline
0.425& &(2/1)&liquid&liquid&liquid&--&--&--&--\\
 & &0.131&--&--&--&--&--&--&--\\
 & &4.986&4.947&4.920&4.907&6.750&6.785&6.786&6.765\\
\hline
0.400& &liquid&liquid&liquid&--&--&--&--&--\\
 & &--&--&--&--&--&--&--&--\\
 & &4.831&4.806&4.796&6.569&6.619&6.629&6.614&6.584\\
\end{tabular}
\end{table}

\end{document}